\documentclass[aip,reprint]{revtex4-1}

\draft % marks overfull lines with a black rule on the right

\usepackage{amsmath,graphicx}

\begin{document}

%\preprint{}

\title{Analytical model of the magnetic field enhancement at pits on the surface of superconducting accelerating cavity} %Title of paper



\author{Takayuki Kubo}
\email[]{kubotaka@post.kek.jp}
%\homepage[]{}
%\thanks{}
\affiliation{KEK, High Energy Accelerator Research Organization, 1-1 Oho, Tsukuba, Ibaraki 305-0801 Japan}


\date{\today}

\begin{abstract}
A simple model of the magnetic field enhancement at pits on the surface of superconducting accelerating cavity is proposed. 
The model consists of a two-dimensional pit with a slope angle, depth, width, and radius of round edge. 
An analytical formula that describes the magnetic field enhancement factor of the model is derived.  
The formula is given as a function of a slope angle and a ratio of half a width to a round-edge radius. 
Agreements between the formula and numerical calculations are also demonstrated. 
Using the formula, the field vortices start to penetrate can be evaluated for a given geometry of pit. 
\end{abstract}

\pacs{}% insert suggested PACS numbers in braces on next line

\maketitle %\maketitle must follow title, authors, abstract and \pacs

%%%%%%%%%%%%%%%%%%%%%%%%%%%%%%%%%%%%%%%%%%%%%%%%%%
%%%%%%%%%%%%%%%%%%%%%%%%%%%%%%%%%%%%%%%%%%%%%%%%%%
\section{Introduction}\label{section:introduction}
%%%%%%%%%%%%%%%%%%%%%%%%%%%%%%%%%%%%%%%%%%%%%%%%%%
%%%%%%%%%%%%%%%%%%%%%%%%%%%%%%%%%%%%%%%%%%%%%%%%%%

After the pioneering study of the superconducting radio-frequency (SRF) cavity for its application to accelerator in the 1960's, 
tremendous efforts to improve their performances have been made~\cite{hassan}.  
The maximum accelerating gradients have been increasing 
from a few ${\rm MV/m}$ in pioneer days to over $35\,{\rm MV/m}$ of TESLA type $1.3\,{\rm GHz}$ 9-cell cavities made of niobium~\cite{TDR}.  
Further a high gradient comparable to the theoretical limit imposed by the superheating field of the bulk niobium, however, 
is difficult to achieve even today, because of quenches at rather low gradients.

Causes of quenches have been identified at many experiments by use of the optical inspection technique~\cite{iwashita} combined with the temperature and X-ray mapping systems~\cite{yamamoto,champion}. 
According to experiments, pits on inner surfaces of SRF cavities are thought to be one of main causes of quenches. 
In fact removals of suspicious pits by the local grind~\cite{grind}, laser re-melting~\cite{laser} and electron beam re-melting technique~\cite{electronremelting} improve maximum accelerating gradients of cavities.

A quench by a pit is attributed to a locally enhanced magnetic field, $\beta H_0$, exceeding the superheating field, 
where $H_0$ is a surface magnetic field of cavity, and $\beta$ is a magnetic field enhancement (MFE) factor due to a geometry of pit. 
To reveal the relation between the MFE factor and the geometry of pit is of high interest not only to a current SRF technology based on the niobium cavity, 
but also to future technologies such as the cavity with the multilayered structure of superconductors~\cite{gurevich, kis}, because  
quenches by a locally enhanced magnetic field at pits remain a concern 
as long as the SRF cavity is used at high magnetic fields comparable to the critical magnetic field of the material. 
The important contribution to this challenge is given by V. Shemelin and H. Padamsee~\cite{shemelin}. 
They demonstrated the geometrical dependence of the MFE factor at a well-like pit by numerical calculations, 
and proposed the simple relation $\beta \propto (R/r_e)^{1/3}$, where $R$ is half a width of the well and $r_e$ is a radius of round edge. 
Based on this study several experiments and numerical studies using well-like pits on surfaces of cavities have been carried out~\cite{xie}.    
Except for artificial pits, however, those found in SRF cavities are usually not well-like, but have gentle slopes~\cite{watanabe, ge}. 
The relation given above can not be applicable to such pits with arbitrary slope angles.  
A more general formula that can describe the MFE factor of pits with arbitrary slope angles is needed.

In this paper, an analytical model of the MFE factor at pits is proposed. 
The model consists of a two-dimensional pit with a slope angle, depth, width, and radius of round edge. 
In Sec.~\ref{Formulation}, the derivation of the MFE factor of this model, which is equal to solving the Maxwell equations of a two-dimensional magnetostatics in the vacuum, 
is reduced to the problem of finding an appropriate complex function called the complex potential. 
Then, in Sec.~\ref{sharp}, the complex potential and the resultant MFE factor for the model with an infinitesimal round-edge radius is derived analytically by the method of conformal mapping, 
from which, in Sec.~\ref{round}, that with a finite round-edge radius is derived by a polynomial extrapolation. 
Agreements between the formula and numerical calculations are also demonstrated. 
In Sec.~\ref{discussion}, by using the formula, the field vortices start to penetrate is evaluated.  
A summary is given in Sec.~\ref{summary}.

%%%%%%%%%%%%%%%%%%%%%%%%%%%%%%%%%%%%%%%%%%%%%%%%%%%%%%%%%%%%%%%%%
%%%%%%%%%%%%%%%%%%%%%%%%%%%%%%%%%%%%%%%%%%%%%%%%%%%%%%%%%%%%%%%%%
\section{Model}%\label{section:model}
%%%%%%%%%%%%%%%%%%%%%%%%%%%%%%%%%%%%%%%%%%%%%%%%%%%%%%%%%%%%%%%%%
%%%%%%%%%%%%%%%%%%%%%%%%%%%%%%%%%%%%%%%%%%%%%%%%%%%%%%%%%%%%%%%%%

\begin{figure}[tb]
   \begin{center}
   \includegraphics[width=0.8\linewidth]{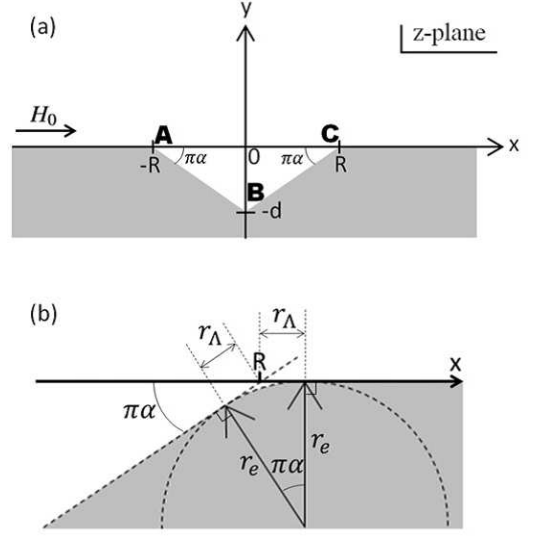}
   \end{center}\vspace{-0.6cm}
   \caption{
(a) A two-dimensional pit with a triangular section and (b) an enlarged image of the edge.  
A gray region corresponds to a superconductor in the Meissner state,  
$\pi \alpha\, (0< \alpha <1/2)$ is a slope angle, $d$ is a depth, $R=d/\tan\pi\alpha$ is half a width of the open mouth, 
and $r_e$ is a radius of the round edge. 
A length $r_{\Lambda}= r_e(1-\cos\pi\alpha)/\sin\pi\alpha$ is a distance between $z=R$ and an end of the arc. 
A surface magnetic field far from the pit is given by $(H_x,H_y)=(H_0,\, 0)$.  
   }\label{figure1}
\end{figure}

Pits found on inner surfaces of SRF cavities have various slope angles, widths, depths and curvature radii of edges. 
A minimum model with such parameters is a two-dimensional pit with a triangular section shown in Fig.~\ref{figure1},  
where $\pi \alpha$ ($0<\alpha < 1/2$) is a slope angle, $d$ is a depth, $R=d/\tan\pi\alpha$ is half a width of the open mouth, 
and $r_e$ is a radius of the arc of the round edge, respectively. 
The magnetic field far from the pit is given by $(H_x,H_y)=(H_0,0)$, which is uniform and parallel to the surface of the superconductor.

%%%%%%%%%%%%%%%%%%%%%%%%
\subsection{Formulation}\label{Formulation}
%%%%%%%%%%%%%%%%%%%%%%%%

The Maxwell equations for a two-dimensional magnetostatics problem in the vacuum are given by ${\rm rot}\,{\bf H}={\bf 0}$ and ${\rm div}\,{\bf B}=0$, 
where ${\bf H}$ is a magnetic field, ${\bf B}=\mu_0{\bf H}$ is a magnetic flux density, and $\mu_0$ is the magnetic permeability of the vacuum.  
By introducing a magnetic scalar potential $\phi(x,y)$, the problem is reduced to a Laplace equation $\triangle^{(2)} \phi(x,y)=0$, 
where $\triangle^{(2)} = \partial^2/\partial x^2 + \partial^2/\partial y^2$.   
The magnetic field is given by ${\bf H}=-{\rm grad}\, \phi = (-\partial \phi/\partial x, -\partial \phi/\partial y, 0)$. 
On the other hand, by introducing a vector potential ${\bf A}=(0, 0, $ $-\mu_0\, \psi(x,y))$, where $\psi$ is a real function, 
the problem is reduced to a Laplace equation $\triangle^{(2)} \psi(x,y)=0$.  
The magnetic field is given by ${\bf H} =\mu_0^{-1}\,{\rm rot}\,{\bf A}=(-\partial \psi/\partial y, \partial \psi/\partial x, 0)$. 
Since both the two approaches should lead the same magnetic field, 
\begin{eqnarray}
H_x = -\frac{\partial \phi}{\partial x} = -\frac{\partial \psi}{\partial y} \,, \hspace{1cm} 
H_y = -\frac{\partial \phi}{\partial y} = \frac{\partial \psi}{\partial x} \,,\label{eq:CR}
\end{eqnarray}
are derived, which are the Cauchy-Riemann conditions.  
Thus a function 
\begin{eqnarray}
\Phi(z) = \phi(x,y) + i\psi(x,y)\,,
\end{eqnarray}
is an holomorphic function of a complex variable $z=x+iy$, which is called the complex potential.  
If a complex potential $\Phi(z)$ is given, components of the magnetic field are derived from 
\begin{eqnarray}
H_x - i H_y 
 = - \frac{\partial \phi}{\partial x} + i\frac{\partial \phi}{\partial y} 
 = -\frac{\partial \phi}{\partial x} - i\frac{\partial \psi}{\partial x} 
 = -\Phi'(z)  
\,,  \label{eq:Hx-iHy1}
\end{eqnarray}
where the property of the holomorphic function, $\Phi'(z)=\partial \phi/\partial x + i\partial \psi/\partial x$, is used. 
Then the MFE factor, which is defined by a ratio of an enhanced magnetic field to a surface magnetic field far from a pit, is given by
\begin{eqnarray}
\beta = \frac{|H_x - i H_y|}{H_0} = \frac{|\Phi'(z)|}{H_0} \,.  \label{eq:beta}
\end{eqnarray}
Thus the two-dimensional magnetostatics problem is reduced to a problem of finding a complex potential $\Phi(z)$.

%

%%%%%%%%%%%%%%%%%%%%%%%%%%%%%%%%%%%%%%%%%%%%%%%%%%%%%%%%%%
\subsection{Pit with sharp edges ($r_e = 0$)}\label{sharp}
%%%%%%%%%%%%%%%%%%%%%%%%%%%%%%%%%%%%%%%%%%%%%%%%%%%%%%%%%%

%
\begin{figure}[t]
   \begin{center}
   \includegraphics[width=0.8\linewidth]{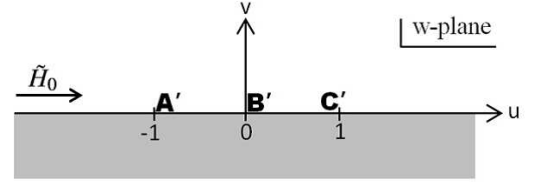}
   \end{center}\vspace{-0.6cm}
   \caption{
   A superconductor surface without a pit. 
A gray region corresponds to a superconductor in the Meissner state.  
A surface magnetic field is assumed to be $(H_u,H_v)=(\tilde{H}_0,0)$. 
   }\label{figure2}
\end{figure}

Let us consider the model with $r_e = 0$ for simplicity, which corresponds to a pit with sharp edges. 
In this case, the complex potential $\Phi(z)$ and the resultant MFE factor are easily derived from a complex potential $\tilde{\Phi}(w)$ 
on a complex $w$-plane shown in Fig.~\ref{figure2} through a conformal mapping $z=F(w)$, 
by which orthogonal sets of field lines in the $w$-plane are transformed into those in the $z$-plane. 
The map $F(w)$ is given by the Schwarz-Christoffel transformation~\cite{laura, miller}: 
\begin{eqnarray}
\!\!\!\!F(w)=K_1 \int_0^w \!\! f(w)  dw + K_2 \, , \hspace{0.3cm} 
f(w) = \Bigl( \frac{w^2-1}{w^2} \Bigr)^{\alpha} . \label{eq:SC}
\end{eqnarray}
Constants $K_1$ and $K_2$ are determined by conditions that B' and C' in the $w$-plane are mapped into B and C in the $z$-plane, respectively,  
and are given by $K_1=\sqrt{\pi}R/\{ \alpha \cos{\pi\alpha} \, \Gamma(\alpha)\Gamma(\frac{1}{2}-\alpha)\} $ and $K_2=-id$.   
The complex potential on the $w$-plane is given by 
\begin{eqnarray}
\tilde{\Phi}(w) = -\tilde{H}_0 w \,,  \label{eq:tildePhi}
\end{eqnarray}
where $\tilde{H}_0 \equiv K_1 H_0$.   
It is easy to confirm that components of the maganetic field on the $w$-plane is given by $H_u - iH_v = -\tilde{\Phi}'(w) = \tilde{H}_0$.   
Thus, from Eq.~(\ref{eq:SC}) and (\ref{eq:tildePhi}), the complex potential on the $z$-plane is given by
\begin{eqnarray}
\Phi(z) = \tilde{\Phi}(F^{-1}(z)) = -\tilde{H}_0 F^{-1}(z)\,, \label{eq:Phiz}   
\end{eqnarray}
where $F^{-1}$ is a inverse function of $F$. 
Then the MFE factor of the model are derived by substituting Eq.~(\ref{eq:Phiz}) into the general formula given by Eq.~(\ref{eq:beta}):  
\begin{eqnarray}
\!\!\!\!\!\!\beta  
= K_1 \biggl| \frac{dF^{-1}(z)}{dz} \biggr|
= \frac{K_1}{|F'(w)|} 
= \frac{1}{|f(w)|} 
= \biggl| \frac{w^2}{w^2-1} \biggr|^{\alpha} ,
\label{eq:beta2}
\end{eqnarray}
where $dF^{-1}/dz = dw/dz= (dz/dw)^{-1} = (dF/dw)^{-1}$ is used. 
As is obvious from the above, the magnetic field is not enhanced ($\beta \to 1$) at a domain far from the pit ($w \to \infty$), 
vanishes ($\beta =0$) at the concave corner B ($w = 0$), and  
diverges ($\beta \to \infty$) at the convex corners A and C ($w\to \pm 1$).

Let us look at the vicinity of the convex corner C in detail. 
In order to evaluate the MFE factor for an arbitrary $z$,  
the function $f(w)$ in Eq.~(\ref{eq:beta2}) is required to be expressed in terms of $z$.   
While no closed form of $f(F^{-1}(z))$ exist, that of an approximate expression can be derived as follows~\cite{miller}. 
Since a domain in the vicinity of C' on the $w$-plane can be written as $w=1+\epsilon$ with $|\epsilon| \ll 1$,  
the function $f(w)=(w^2-1)^{\alpha}w^{-2\alpha}$ can be approximated as $f(w) \simeq 2^{\alpha} \epsilon^{\alpha}$. 
Then the integral in Eq.~(\ref{eq:SC}) is computed as $z=F(w)=R + 2^{\alpha} K_1 \epsilon^{\alpha+1}/ (\alpha +1)$, or  
\begin{eqnarray}
f(w) \simeq  2^{\alpha} \epsilon^{\alpha} = \Bigl[ \frac{2(\alpha +1)}{K_1} ( z- R) \Bigr]^{\gamma} , 
\hspace{0.3cm} 
\gamma \equiv \frac{\alpha}{1+\alpha} \,.  \label{eq:approxfw}
\end{eqnarray}
Substituing Eq.~(\ref{eq:approxfw}) into Eq.~(\ref{eq:beta2}), we find 
\begin{eqnarray}
\beta(r) = \Bigl[  \frac{\sqrt{\pi} }{2\alpha(\alpha +1) \cos{\pi\alpha} \, \Gamma(\alpha)\Gamma(\frac{1}{2}-\alpha)} \frac{R}{r} \Bigr]^\gamma \,. 
\label{eq:beta3}
\end{eqnarray}
where $r\equiv |z - R|$ is a distance from the corner C. 
It should be noted that the range in application of Eq.~(\ref{eq:beta3}) is limited by the assumption $\epsilon \ll 1$, and is given by
\begin{eqnarray}\label{eq:range}
r \ll r_{\epsilon} \equiv \frac{2^\alpha \sqrt{\pi} R}{\alpha(\alpha +1) \cos{\pi\alpha} \, \Gamma(\alpha)\Gamma(\frac{1}{2}-\alpha)} \, . 
\end{eqnarray}
The MFE factor near the corner A can also be described by Eq.~(\ref{eq:beta3}), 
by replacing $r=|z-R|$ with $r=|z+R|$, a distance from the corner A.

%%%%%%%%%%%%%%%%%%%%%%%%%%%%%%%%%%%%%%%%%%%%%%%
\subsection{Pit with round edges ($r_e \ne 0$)}\label{round}
%%%%%%%%%%%%%%%%%%%%%%%%%%%%%%%%%%%%%%%%%%%%%%%

Let us proceed to the model with a finite $r_e$. 
As is obvious from Fig.~\ref{figure1}(b), at a domain $r \ge r_{\Lambda} \equiv r_e(1-\cos\pi\alpha)/\sin\pi\alpha$, 
the pit has the same geometry as that with sharp edges. 
Thus the MFE factor $\beta(r)$ of the model with a finite $r_e$ is also described by Eq.~(\ref{eq:beta3}) at $r_{\Lambda} \le r \ll r_{\epsilon}$, 
where $r_{\epsilon}$ is given by Eq.~(\ref{eq:range}).  
Note that $r_{\Lambda} \ll r_{\epsilon}$, namely,  
\begin{eqnarray}\label{eq:rangeR/re}
\frac{R}{r_e} \gg \lambda (\alpha) \equiv \frac{\alpha(\alpha +1) (1-\cos{\pi\alpha}) \, \Gamma(\alpha)\Gamma(\frac{1}{2}-\alpha)}{2^\alpha \sqrt{\pi} \tan\pi\alpha} \, , 
\end{eqnarray}
is assumed here.

The MFE factor $\beta(r)$ at $r \le r_{\Lambda}$ can be estimated from that at $r \ge r_{\Lambda}$ by a polynomial extrapolation,   
\begin{eqnarray}
\beta(r) = \sum_{k=0}^n a_k (r-r_{\Lambda})^k \,,
\end{eqnarray}
where $a_k$ is a real coefficient and $n$ is a degree of the polynomial. 
Suppose the polynomial satisfies following conditions: 
continuities of $\beta$, $d\beta/dr$, $\dots,\,d^{n-1}\beta/dr^{n-1}$ at $r = r_{\Lambda}$, and $d\beta/dr \simeq 0$ at $r \ll r_{\Lambda}$. 
Due to the last condition, the function $\beta(r)$ has its maximum value at $r\ll r_{\Lambda}$, the tip of the round edge. 
Then the coefficients are given by 
\begin{eqnarray}\label{eq:coefficients}
\!\!\!\!\!a_k =
  \begin{cases}
    \beta(r_{\Lambda})                                                                                  & (k=0) \,, \\
    (-1)^k \gamma (1+\gamma) \cdots (k-1+\gamma) \frac{\beta(r_{\Lambda})}{k! r_{\Lambda}^k}              & \hspace{-1.1cm}(1\le k \le n-1) \,, \\
    (-1)^{k+1} \gamma (2+\gamma) \cdots (n-1+\gamma) \frac{(n-1) \beta(r_{\Lambda})}{n! r_{\Lambda}^n}  & (k=n) \,, 
  \end{cases} \nonumber
\end{eqnarray}
where $\beta(r_{\Lambda})$ can be evaluated by Eq.~(\ref{eq:beta3}) when Eq.~(\ref{eq:rangeR/re}) is satisfied. 
Thus the maximum value of the MFE factor $\beta^* \simeq \beta(r)|_{r \ll r_{\Lambda}}$ 
is given by  
\begin{eqnarray}
\beta^* \,\, 
&\simeq& \sum_{k=0}^n a_k (-1)^k r_{\Lambda}^k \nonumber \\
&=& \frac{\beta(r_{\Lambda})}{(n-1)!}\Bigl( 1+\frac{\gamma}{n}\Bigr) (2+\gamma) \cdots (n-1+\gamma) \nonumber \\
&=& P(\alpha) \, \biggl( \frac{R}{r_e} \biggr)^\gamma
\,, \label{eq:betaI}
\end{eqnarray}
where the coefficient $P(\alpha)$ is given by 
\begin{eqnarray} \label{eq:p}
\!\!\!\!\!\!\!\!\!P(\alpha) 
&=&\frac{1}{(n-1)!}\Bigl( 1+\frac{\gamma}{n}\Bigr) (2+\gamma) \cdots (n-1+\gamma)   \nonumber\\ 
&\times &
\biggl[  
\frac{\sqrt{\pi} \tan\pi\alpha }
     {2\alpha(\alpha +1)(1\!-\!\cos\pi\alpha) \Gamma(\alpha)\Gamma(\frac{1}{2}\!-\!\alpha)} 
\biggr]^\gamma .  
\end{eqnarray}
\begin{figure}[tb]
   \begin{center}
   \includegraphics[width=0.7\linewidth]{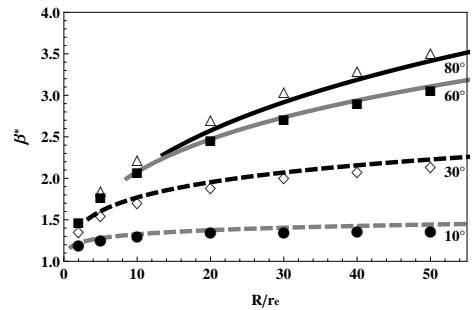}
   \end{center}\vspace{-0.6cm}
\caption{
Comparisons between the analytical formula given by Eq.~(\ref{eq:betaI}) and numerical calculations by POISSON/SUPERFISH. 
The four curves show Eq.~(\ref{eq:betaI}) as functions of $R/r_e$ in ranges $R/r_e \ge 10\lambda(\alpha)$ for Eq.~(\ref{eq:rangeR/re}).  
The black line, gray line, black dashed line and gray dashed line correspond to 
slope angles $80^{\circ}$, $60^{\circ}$, $30^{\circ}$, and $10^{\circ}$ ($\alpha = 4/9$, $1/3$, $1/6$, and $1/18$), respectively. 
The four types of symbols show the maximum MFE factors for given pits calculated by POISSON/SUPERFISH~\cite{poisson}, 
where the triangles, filled squares, rhombi, and filled circles correspond to 
slope angles $80^{\circ}$, $60^{\circ}$, $30^{\circ}$, and $10^{\circ}$, respectively. 
}\label{figure3}
\end{figure}
Fig.~\ref{figure3} makes comparisons between Eq.~(\ref{eq:betaI}) for $n=6$ and numerical calculations by POISSON/SUPERFISH~\cite{poisson},  
where good agreements between the two approaches are seen. 
Even though $\beta^*$ is proportional to $(R/r_e)^{\gamma}$ independently of a choice of the degree of polynomial $n$, 
other choice of $n$ does not give such good agreements, because of too large or too small $P(\alpha)$. 
It is worth emphasizing that Eq.~(\ref{eq:betaI}) is reduced to $\beta^* \propto (R/r_e)^{1/3}$ when $\alpha \to 1/2$, which corresponds to the known result for the well-like pit~\cite{shemelin}.

%%%%%%%%%%%%%%%%%%%%%%%%%%%%%%%%%%%%%%%%%%%%%%%%%%%%%%%%%%%%%%
%%%%%%%%%%%%%%%%%%%%%%%%%%%%%%%%%%%%%%%%%%%%%%%%%%%%%%%%%%%%%%
\section{Discussion}\label{discussion}
%%%%%%%%%%%%%%%%%%%%%%%%%%%%%%%%%%%%%%%%%%%%%%%%%%%%%%%%%%%%%%
%%%%%%%%%%%%%%%%%%%%%%%%%%%%%%%%%%%%%%%%%%%%%%%%%%%%%%%%%%%%%%

\begin{figure}[tb]
   \begin{center}
   \includegraphics[width=0.8\linewidth]{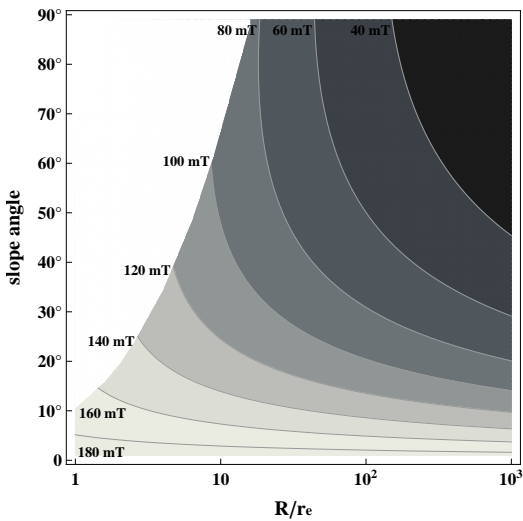}
   \end{center}\vspace{-0.6cm}
   \caption{
A contour plot of $\mu_0 H_{\rm peak}^{\rm max}$.  
The abscissa and the ordinate represent the ratio $R/r_e$ and the slope angle $\pi \alpha$, respectively, which specify the geometry of a pit.  
The white region at the upper left corresponds to a region $R/r_e < 10 \lambda(\alpha)$, 
where our formulae ceases to be a good approximation. 
   }\label{figure4}
\end{figure}

Using the formula given by Eq.~(\ref{eq:betaI}), 
the field vortices start to penetrate can be evaluated by a geometry of pit characterized by a slope angle $\alpha$ and a ratio $R/r_e$. 
Suppose a pit exist on an equator of a superconducting accelerating cavity. 
Then a peak magnetic field of the cavity, $H_{\rm peak}$, is maximally enhanced at the edge of the pit: $\beta^* H_{\rm peak}$. 
Provided that vortices start to penetrate into the superconductor at $H_v$, 
the maximum peak magnetic field in the absence of the vortex penetration, $H_{\rm peak}^{\rm max}$, can be estimated as $\beta^* H_{\rm peak}^{\rm max} = H_v$, or,  
\begin{eqnarray}
\mu_0 H_{\rm peak}^{\rm max} = \frac{\mu_0 H_v}{P(\alpha)} \biggl( \frac{r_e}{R} \biggr)^{\gamma} \,.  \label{eq:H0max}
\end{eqnarray}
Fig.(\ref{figure4}) shows the contour plot of $\mu_0 H_{\rm peak}^{\rm max}$, where $\mu_0 H_v = 200\,{\rm mT}$ is assumed~\cite{valles}. 
It should be noted that if a particular shape of cavity is assumed, Eq.~(\ref{eq:H0max}) is reduced to the formula that describes the accelerating field at which vortices start to penetrate:  
$E_{\rm acc}^{\rm max} = g^{-1} \mu_0 H_{\rm peak}^{\rm max}$, 
where $g$ is a ratio of the peak magnetic field to the accelerating field for a given cavity. 
It is also easy to draw the similar contour plot for cavities with a different $\mu_0 H_v$ such as a multi-layered structure of the superconductors~\cite{srf13kis}. 
A comparison between the formula and experimental results is a future challenge.

%%%%%%%%%%%%%%%%%%%%%%%%%%%%%%%%%%%%%%%%%%%%%%%%%%%%%%%%%%%%%%
%%%%%%%%%%%%%%%%%%%%%%%%%%%%%%%%%%%%%%%%%%%%%%%%%%%%%%%%%%%%%%
\section{Summary}\label{summary}
%%%%%%%%%%%%%%%%%%%%%%%%%%%%%%%%%%%%%%%%%%%%%%%%%%%%%%%%%%%%%%
%%%%%%%%%%%%%%%%%%%%%%%%%%%%%%%%%%%%%%%%%%%%%%%%%%%%%%%%%%%%%%

A simple model of the magnetic field enhancement at pits on the surface of superconducting accelerating cavity is proposed. 
The model consists of a two-dimensional pit with 
a slope angle $\pi \alpha$ ($0<\alpha < 1/2$), depth $d$, width $R$, and radius of round edge $r_e$.  
A formula that describes the magnetic field enhancement factor for the model was derived.  
The formula is given by $\beta^* = P(\alpha) (R/r_e)^{\gamma}$, 
where $\gamma$ is defined by $\gamma=\alpha/(1+\alpha)$, and $P(\alpha)$ is a coefficient depending on a slope angle $\alpha$.  
By using the formula, the field vortices start to penetrate can be evaluated for a given geometry of pit.

%%%%%%%%%%%%%%%%%%%%%%%%%%%%%
%%%%%%%%%%%%%%%%%%%%%%%%%%%%%
%\section*{Acknowledgements}
%%%%%%%%%%%%%%%%%%%%%%%%%%%%%
%%%%%%%%%%%%%%%%%%%%%%%%%%%%%

%%%%%%%%%%%%%%%%%%%%%%%%%%%
%%%%%%%%%%%%%%%%%%%%%%%%%%%

\bibliography{your-bib-file}

\end{document}